\documentclass[twocolumn,showpacs,preprintnumbers,prl,nofootinbib]{revtex4-1}
\usepackage{graphicx,epsfig,amsmath,amssymb,bm,slashed,wasysym}
\usepackage{color}
\makeatletter
\DeclareRobustCommand*{\bfseries}{%
  \not@math@alphabet\bfseries\mathbf
  \fontseries\bfdefault\selectfont
  \boldmath
}
\makeatother

\begin{document}

\preprint{} 

\title{On minimal $Z^\prime$ explanations of the $B \to K^\ast \mu^+ \mu^-$ anomaly}

\author{Rhorry Gauld$^a$, Florian Goertz$^b$ and Ulrich Haisch$^c$} 

\affiliation{${}^a$\,Department of Physics, University of Oxford, 
OX1 3PN Oxford, United Kingdom\\
${}^b$\,Institute for Theoretical Physics, ETH Zurich, 8093 Zurich, Switzerland\\
${}^c$\,Rudolf Peierls Centre for Theoretical Physics, University of Oxford, 
OX1 3PN Oxford, United Kingdom}

\date{\today}


\begin{abstract}
Recently  LHCb has announced a discrepancy of $3.7 \sigma$  in one of the theoretically clean observables accessible through studies of angular correlations in $B \to K^\ast \mu^+ \mu^-$.  We point out that in the most  minimal $Z^\prime$ setup that can address this anomaly there is a model-independent triple-correlation between new physics (NP) in $B \to K^\ast \mu^+ \mu^-$, $B_s$--$\bar B_s$ mixing, and non-unitarity of the  quark-mixing matrix. This triple-correlation can be cast into a simple analytic formula that relates the NP contribution $\Delta C_9$ to the Wilson coefficient of the semileptonic vector operator to a shift in the mass difference $\Delta M_{B_s}$ and a violation of $ |V_{ud}|^2+ |V_{us}|^2+ |V_{ub}|^2 = 1$. In contrast to the individual observables the found relation depends only logarithmically on the $Z^\prime$ mass.  We show that that our findings  allow for useful future tests of the pattern of NP suggested by the $B \to K^\ast \mu^+ \mu^-$ anomaly.  
\end{abstract}

\maketitle

\subsubsection{Introduction}

The latest LHCb measurements of the decay distributions in $B \to K^\ast \mu^+ \mu^-$ display several  deviations from the standard model (SM) predictions~\cite{Serra,Aaij:2013qta}. With $3.7 \sigma$  the most significant discrepancy arises in the variable $P_5^\prime$~\cite{DescotesGenon:2012zf} (the analogue of $S_5$ in~\cite{Altmannshofer:2008dz}), which combines theoretical and experimental benefits, while retaining a high sensitivity to NP effects in $b \to s \gamma, \ell^+ \ell^-$. Further LHCb studies combined with a critical assessment of theoretical errors (see in particular~\cite{Jager:2012uw}) will be necessary to clarify whether the observed deviations are a real sign of NP or simply flukes. 

Shortly after  the LHCb announcement, the new results have  been combined  with existing data on other rare and radiative $b \to s$ modes into global fits~\cite{Descotes-Genon:2013wba,Altmannshofer:2013foa}. The most surprising outcome of the analysis \cite{Descotes-Genon:2013wba} is that  the whole pattern of deviations seen by LHCb can be explained by adding a single effective interaction of the form\footnote{The analysis~\cite{Altmannshofer:2013foa} finds that both a left-handed and right-handed semileptonic vector operator are needed to obtain a good fit to the existing $b \to s$ data. Note that  given the left-handedness of the $W$ boson the inclusion of a right-handed $\bar s Z^\prime b$ interaction does not induce a non-unitarity of the quark-mixing matrix, if fermion masses are neglected.} 
\begin{equation} \label{Eq1}
 {\cal L}_{\rm eff} = \frac{4 G_F}{\sqrt{2}} \, \frac{\alpha}{4 \pi} \, V_{ts}^\ast V_{tb} \, C_9  \left (\bar s \gamma_\alpha P_L b \right ) (\bar \mu \gamma^\alpha \mu) + {\rm h.c.}\,, 
\end{equation}
to the SM Lagrangian if the NP effects $\Delta C_9$ in the Wilson coefficient $C_9 = C_9^{\rm SM} +  \Delta C_9 \simeq 4.1 +  \Delta C_9$ \cite{Bobeth:2003at} are large and destructive~\cite{Descotes-Genon:2013wba}  
\begin{equation} \label{Eq2}
\Delta C_9 \sim -1.5 \,. 
\end{equation}
This solution  is intriguing not only because it is pure and simple but also because it is  highly non-standard and cannot be obtained -- at least to our knowledge -- in the most common NP models such as supersymmetry, extra dimensions or partial compositeness (the study \cite{Altmannshofer:2013foa} confirms this naive expectation). 

An obvious though ad hoc way to obtain~(\ref{Eq2}) is to postulate the existence of a new neutral gauge boson (a $Z^\prime$) with TeV-scale mass $M_{Z^\prime}$ and rather particular couplings to fermions~\cite{Descotes-Genon:2013wba}: to avoid disastrous CP-violating contributions to $B_s$--$\bar B_s$ mixing the $Z^\prime$  should couple proportionally to the combination $V_{ts}^\ast V_{tb}$ of Cabibbo-Kobayashi-Maskawa~(CKM) matrix elements to the left-handed $\bar s b$ current; since the $B \to K^\ast \mu^+ \mu^-$ data seem to prefer a vector rather than an axial-vector coupling to the $\bar \mu \mu$ bi-linear, the $Z^\prime$  should  furthermore  feature  left-handed and right-handed  muon  couplings of  close to equal strength. The explicit construction of a realistic $Z^\prime$ model with these properties and its rich phenomenology will be presented elsewhere~\cite{inprep}.

In this letter we  study a phenomenological model that has the above features built in by assumption. We point out that in such a simplified theory there is a model-independent triple-correlation between  $\Delta C_9$, the relative shift $\Delta_{B_s}$ in the mass difference  of the $B_s$-meson system and first-row unitarity violation of the CKM matrix parametrised by $\Delta_{\rm CKM}$. While the correlation between the variables $\Delta C_9$ and $\Delta_{B_s}$ is well known their connection to $\Delta_{\rm CKM}$ has, as far as we are aware, not been discussed in the literature. We show that by means of the triple-correlation it is possible to write the NP contribution $\Delta C_9$ as a simple  analytic function of the shifts $\Delta_{B_s}$ and $\Delta_{\rm CKM}$. In this way the quadratic sensitivity  of $\Delta C_9$  on the inverse of the mass $M_{Z^\prime}$  is turned into a logarithmic dependence.  This implies not only that an observation of $\Delta C_9 \neq 0$ necessarily leads to a deviation  in $B_s$--$\bar B_s$ mixing ($\Delta_{B_s} \neq 0$) and a violation of CKM unitarity ($\Delta_{\rm CKM} \neq 0$), but also that the pattern of the modifications depends weakly on the NP scale. In the case of minimal $Z^\prime$ models, the consistency of~(\ref{Eq2}) can hence be tested  in a simple manner against theoretically clean quark-flavour observables. We discuss the present status and future prospects of these tests as well as  other cross-checks that can be performed at SuperKEKB. 

\subsubsection{Toy model}

Assuming lepton-flavour universality the $Z^\prime$ interactions described in the introduction take the following form 
\begin{equation} \label{Eq3}
\begin{split}
{\cal L} _{Z^\prime} & \supset \left ( V_{ts}^\ast V_{tb}\, g_L^{bs} \, \bar s \slashed{Z}^\prime P_L b + {\rm h.c.} \right ) \\[2mm] & \phantom{xx}  + \frac{g_V^{\mu}}{2} \sum_{\ell = e,\mu,\tau} \left  ( \bar \ell \slashed{Z}^\prime \ell +  \bar \nu_\ell \slashed{Z}^\prime P_L \nu_\ell \right ) \,, 
\end{split}
\end{equation}
where the coupling constants $g_L^{bs}$ and $g_V^\mu$ are real and $P_L = (1 -\gamma_5)/2$ projects out left-handed fields. Note that  $g_V^{\mu}=g_R^{\mu}+g_L^{\mu}=2g_L^{\mu}$, $g_A^{\mu}=g_R^{\mu}-g_L^{\mu}=0$ and $g_L^{\nu_\ell} = g_L^\ell = g_V^{\mu}/2$, where the final relation is a consequence of  $SU(2)_L$ invariance. Throughout our work we will assume that~(\ref{Eq3}) encodes all relevant $Z^\prime$-fermion interactions.  

\subsubsection{Triple-correlation}

The new-physics correction $\Delta C_9$  to the Wilson coefficient of the semileptonic vector operator in~(\ref{Eq1}) is found by calculating the tree-level $Z^\prime$-exchange contribution to the partonic $b \to s \mu^+ \mu^-$ process. We obtain 
\begin{equation} \label{Eq4}
\Delta C_9 = -\frac{1}{\sqrt{2} G_F} \frac{\pi}{\alpha} \frac{g_L^{bs} \, g_V^{\mu}}{M_{Z^\prime}^2} \,,
\end{equation}
where $G_F \simeq 1.167 \cdot 10^{-5} \, {\rm GeV}^{-2}$ denotes the Fermi constant and $\alpha \simeq 1/128$. 

Tree-level $Z^\prime$ exchange also affects the mass difference $\Delta M_{B_s}$. We find 
\begin{equation} \label{Eq5}
\Delta_{B_s} =  \frac{\Delta M_{B_s}}{\Delta M_{B_s}^{\rm SM}} - 1 = \frac{2 \sqrt{2}}{G_F} \frac{\pi s_w^2}{\alpha}  \frac{\tilde r}{S} \frac{\left ( g_L^{bs} \right)^2 }{M_{Z^\prime}^2} \,,
\end{equation}
where $s_w^2 \simeq 0.23$ denotes the sine of the weak mixing angle and $S \simeq 2.3$ \cite{Inami:1980fz} is  the leading-order SM box contribution. The parameter $\tilde r$ encodes renormalisation group effects and is given by \cite{Buras:2012dp}
\begin{equation} \label{Eq6}
\tilde r \simeq 0.985 \left [ 1 - 0.029 \, \ln \left ( \frac{M_Z^\prime}{1 \, {\rm TeV}} \right ) \right ] \,.
\end{equation}

\begin{figure}[t!]
\begin{center}
\makebox{\includegraphics[width=0.85\columnwidth]{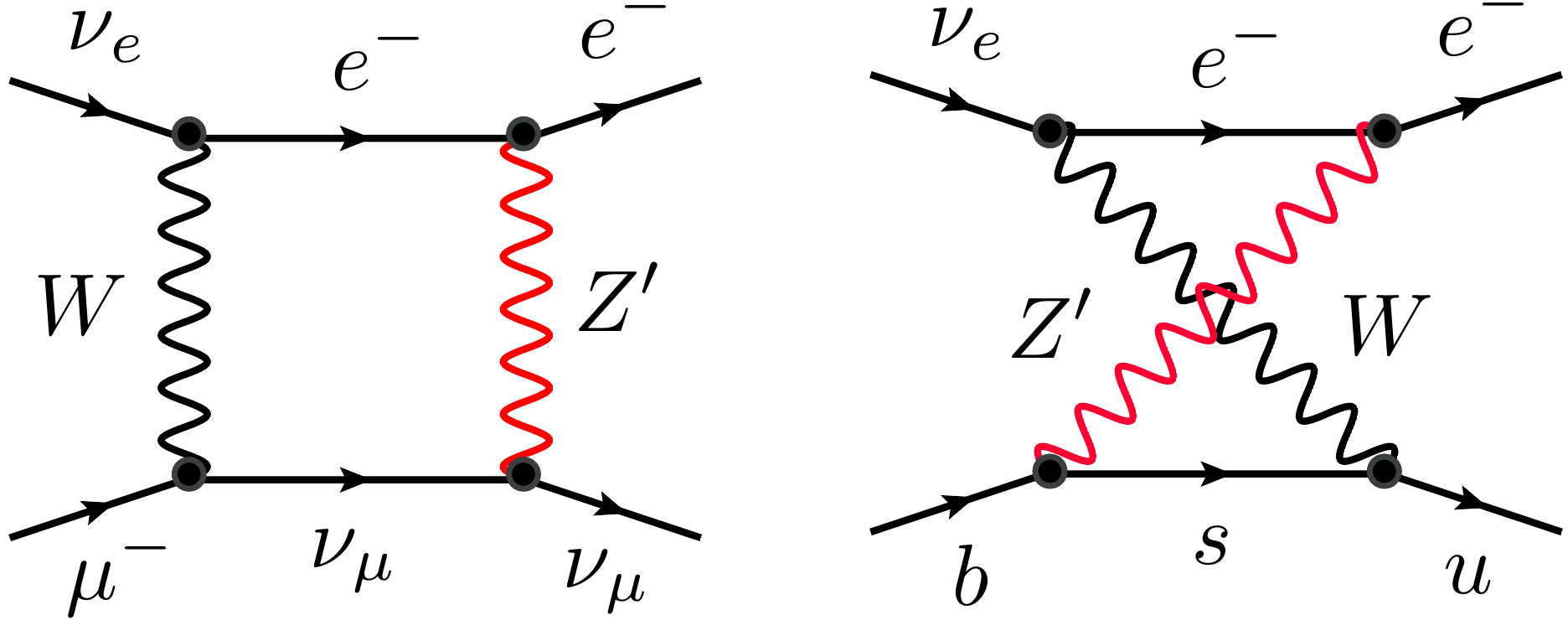}}  
\end{center}
\vspace{-2mm}
\caption{Examples of one-loop box corrections to muon (left) and bottom-quark (right) decays involving $W$ and $Z^\prime$ bosons. For the case of the strange-quark decay the roles of $b$ and $s$ in the right diagram are interchanged.}
\label{Fig1}
\end{figure}

A violation of first-row CKM unitarity is a classic probe of additional neutral gauge bosons \cite{Marciano:1987ja}.  The amount of CKM unitarity violation is determined from  the difference of the one-loop $Z^\prime$ corrections to quark $\beta$-decay amplitudes from which the CKM elements are extracted as well as muon decay which normalises those amplitudes. Examples of  Feynman diagrams relevant in our toy model~(\ref{Eq3}) are shown in Fig.~\ref{Fig1}. Notice that the contributions to $b \to u e^- \bar \nu_e$ ($s \to u e^- \bar \nu_e$) are suppressed relative to  $\mu^- \to e^- \nu_\mu \bar \nu_e$ by $V_{ts}^\ast V_{tb} V_{us} = {\cal O} (\lambda^3) = {\cal O} (1\%)$ $\big($$V_{tb}^\ast V_{ts} V_{ub} = {\cal O} (\lambda^5) = {\cal O} (1 \permil)$$\big)$ with $\lambda \simeq 0.23$ denoting the Cabibbo angle. The flavour-changing contributions to CKM unitarity violation are hence for all practical purposes negligible, and one obtains  \cite{Marciano:1987ja}
\begin{equation} \label{Eq7}
\begin{split}
\Delta_{\rm CKM} & =  \sum_{q=d,s,b} |V_{uq}|^2 -1\\ & = \frac{3}{16 \sqrt{2} G_F} \frac{\alpha}{\pi s_w^2} \frac{\left (  g_V^{\mu}\right )^2 }{M_{Z^\prime}^2} \, \frac{\ln \left ( \frac{M_W^2}{M_{Z^\prime}^2} \right )}{1-  \frac{M_W^2}{M_{Z^\prime}^2} }\,,
\end{split}
\end{equation}
with $M_W \simeq 80.4 \, {\rm GeV}$ denoting the $W$-boson mass. 

The relations~(\ref{Eq5}) and~(\ref{Eq7}) can now be used to eliminate the factor $g_L^{bs} \, g_V^{\mu}/M_{Z^\prime}^2$ entering~(\ref{Eq4}) in favor of $\Delta_{B_s}$ and $\Delta_{\rm CKM}$. Keeping only the leading-logarithmic term in the Taylor expansion of~(\ref{Eq7}) around $M_W^2/M_{Z^\prime}^2 = 0$ $\big($which is an excellent approximation for $M_{Z^\prime} = {\cal O} (1 \, {\rm TeV})$$\big)$, we get 
\begin{equation} \label{Eq8}
\Delta C_9 = -\frac{2\pi}{\sqrt{3} \alpha} \left [ \frac{S}{\tilde r} \frac{|\Delta_{B_s}| |\Delta_{\rm CKM}| }{\ln \left ( \frac{M_{Z^\prime}^2}{M_{W}^2} \right ) } \right ]^{1/2} \,. 
\end{equation}
This is the simple formula advertised in the abstract and the introduction. 

\begin{figure}[t!]
\begin{center}
\makebox{\includegraphics[width=0.85\columnwidth]{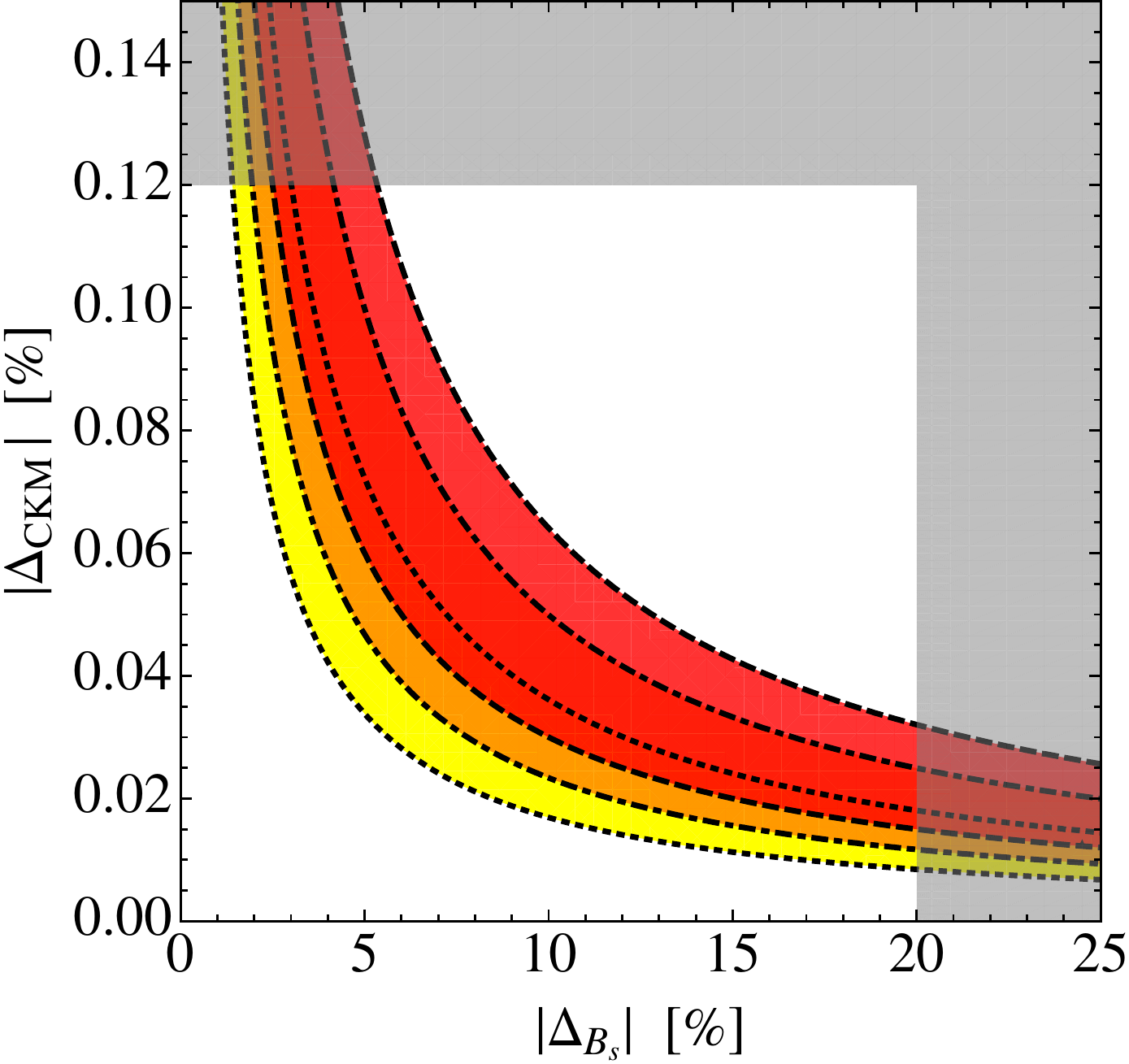}}  

\vspace{2.5mm}

\makebox{\includegraphics[width=0.85\columnwidth]{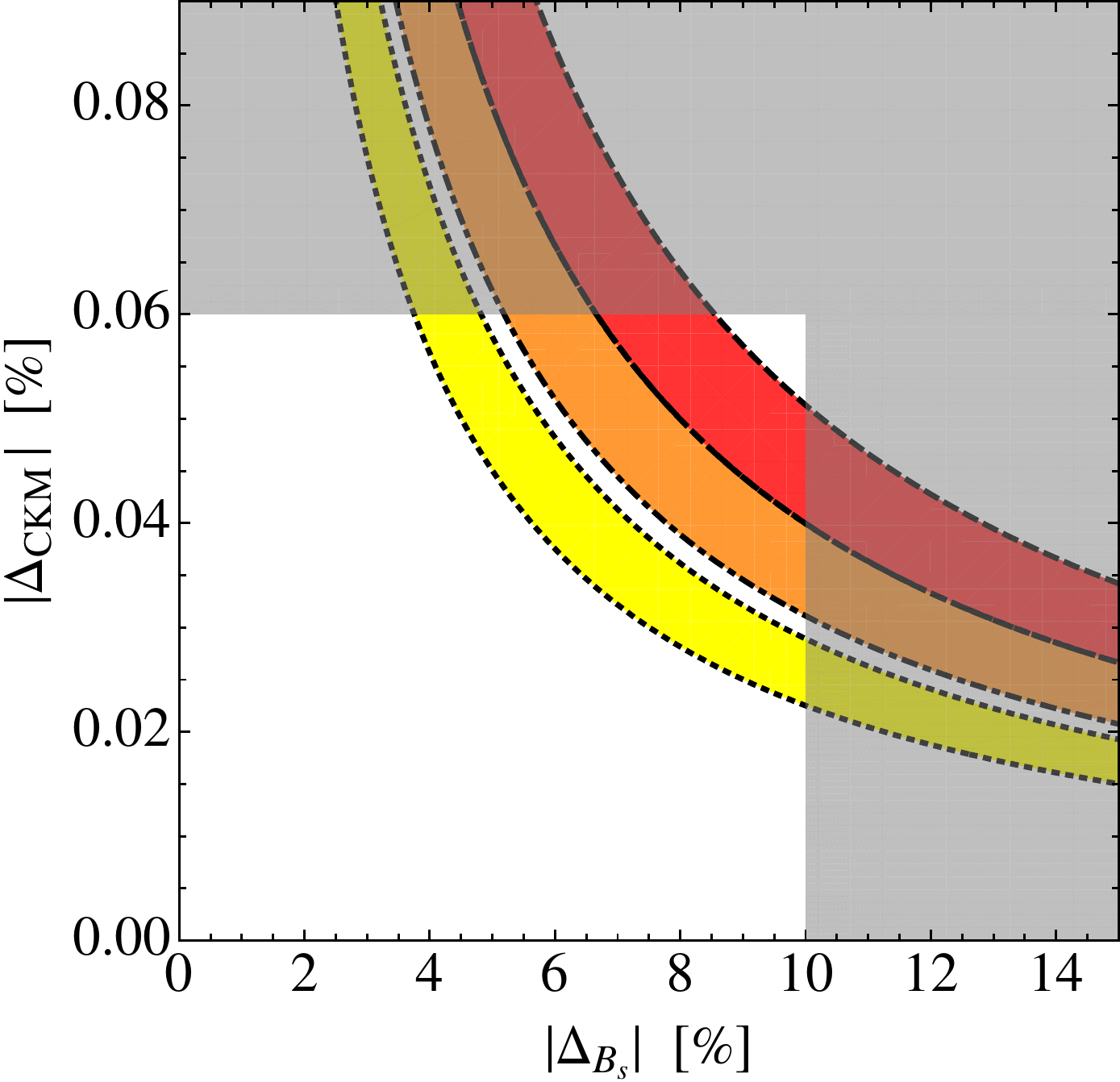}}  
\end{center}
\vspace{-3mm}
\caption{Upper panel: parameter space in the $|\Delta_{B_s}|$--$|\Delta_{\rm CKM}|$ plane favoured by the $B \to K^\ast \mu^+ \mu^-$ anomaly. The gray regions indicate the parameter ranges that are presently disfavoured at 95\% CL. Lower panel: a  possible future projection assuming a confirmation of the $B \to K^\ast \mu^+ \mu^-$ anomaly with improved statistics as well as a reduction of the theory uncertainty in $|\Delta_{B_s}|$ and $|\Delta_{\rm CKM}|$. The yellow, orange and red regions in both plots correspond to $M_{Z^\prime} = 1, 3$ and $10 \, {\rm TeV}$, respectively. Consult the text for further explanations.}
\label{Fig2}
\end{figure}

By means of~(\ref{Eq8}) we can now check the consistency of the preferred fit solution~(\ref{Eq2}) against the non-observation of NP in the mass difference $\Delta M_{B_s}$  \cite{Aaij:2013mpa} and the absence of CKM unitarity violation. At present one has \cite{Descotes-Genon:2013wba}
\begin{equation} \label{Eq9}
\Delta C_9 = [-1.9, -1.3] \,,
\end{equation}
at 68\% confidence level (CL), while \cite{UTFit, Antonelli:2010yf} 
\begin{equation} \label{Eq10}
|\Delta_{B_s} | < 20 \%\,, \qquad |\Delta_{\rm CKM} | < 1.2 \permil \,,
\end{equation}
at 95\% CL.  The resulting constraints are shown in the upper panel of Fig.~\ref{Fig2}. The yellow, orange and red region in the plot corresponds to a $Z^\prime$ mass of $1, 3$ and $10 \, {\rm TeV}$, respectively, while the gray area indicates values of $|\Delta_{B_s}|$ and/or $|\Delta_{\rm CKM}|$ inconsistent with~(\ref{Eq10}).  We see that a large and negative shift  $\Delta C_9$, as suggested by the $B \to K^\ast \mu^+ \mu^-$  anomaly, is consistent in our toy model~(\ref{Eq3}) only if $|\Delta_{B_s}|$ and $|\Delta_{\rm CKM}|$ are both non-zero and obey $\sqrt{\left |\Delta_{B_s} \right | \left  | \Delta_{\rm CKM} \right | } = {\rm const.}$ (for fixed $\Delta C_9$ and $M_{Z^\prime}$).  Interestingly, explaining the anomaly in $B \to K^\ast \mu^+ \mu^-$ with heavier $Z^\prime$ bosons, one faces stronger constraints from $\Delta_{B_s}$ and $\Delta_{\rm CKM}$ due to the appearance of the logarithm $\ln \hspace{0.25mm} (M_{Z^\prime}^2/M_W^2)$ in the denominator of~(\ref{Eq9}). 

To  illustrate the potential of the triple-correlation as a cross-check, we also present results of a possible, though hypothetical, future scenario. We assume that the $B \to K^\ast \mu^+ \mu^-$ anomaly is confirmed with higher significance leading to 
\begin{equation} \label{Eq11}
\Delta C_9 = [-1.7, -1.5] \,,
\end{equation}
and that the theoretical understanding of $\Delta_{B_s}$ and $\Delta_{\rm CKM}$ is improved by a factor of 2, implying 
\begin{equation} \label{Eq12}
|\Delta_{B_s} | < 10 \%\,, \qquad |\Delta_{\rm CKM} | < 0.6 \permil \,.
\end{equation}
In this case we obtain the results shown in the lower panel of Fig.~\ref{Fig2}. It is evident  from the plot that in such a futuristic scenario an explanation of~(\ref{Eq11}) by a minimal $Z^\prime$  with mass of ${\cal O} (5 \, {\rm TeV})$ would become testable. This is interesting because such heavy $Z^\prime$ bosons would very likely escape direct detection at the LHC even at 14 TeV and high luminosity. Future improvements in lattice-QCD determinations of  the $B_s$-meson decay constant $f_{B_s}$, the hadronic parameter $\hat B_{B_s}$ as well as $V_{cb}$, which represent the dominant sources of error in $\Delta M_s^{\rm SM}$ \cite{Lenz:2006hd}, could hence also have a vital impact on $B \to K^\ast \mu^+ \mu^-$ studies. A similar statement applies to improved tests of CKM unitarity that call for better determinations of $|V_{ud}|$ and $|V_{us}|$.  In the long run, extractions of $|V_{ud}|$ by future neutron decay studies \cite{Markisch:2011ik} could play a key role here, since they are, unlike nuclear beta decay,  not limited by the theoretical knowledge of nuclear corrections.

\subsubsection{Other implications}

Constraints on the structure of $\bar s Z^\prime b$ and $\bar \mu Z^\prime \mu$ interactions arise also from the measurements of $B_s \to \mu^+ \mu^-$  by LHCb~\cite{Aaij:2013aka} and CMS~\cite{Chatrchyan:2013bka}. In this context it is important to realise that our phenomenological model predicts ${\rm Br} \hspace{0.25mm} (B_s \to \mu^+ \mu^-) = {\rm Br}  \hspace{0.25mm}  (B_s \to \mu^+ \mu^-)_{\rm SM}$, since the axial-vector coupling between the $Z^\prime$ and muons is set to zero by hand. Finding a notable deviation from the SM in  $B_s \to \mu^+ \mu^-$ would thus imply that the structure of~(\ref{Eq3})  has to be extended by allowing for $g_A^\mu \neq 0$. Such a modification will unavoidably lead to a correlated effect in $B \to K^\ast \mu^+ \mu^-$ that can be probed and constrained  by further LHCb data. Accurate measurements of the variable $P_4^\prime$~\cite{DescotesGenon:2012zf} (or $S_4$ in the notation of~\cite{Altmannshofer:2008dz}) will play a key role in this context~\cite{Quim}.  

Since $SU(2)_L$ invariance requires $g_L^{\nu_\ell} = g_L^\ell$, a deviation like~(\ref{Eq2}) will also affect the $b \to s \nu \bar \nu$ transitions. In the case of the minimal model~(\ref{Eq3}), one arrives at the prediction ($F = K^\ast, K, X_s$)
\begin{equation} \label{Eq14}
\begin{split}
\Delta_{\nu \bar \nu} & = \frac{{\rm Br} \hspace{0.25mm} (B \to F \nu \bar \nu) }{{\rm Br} \hspace{0.25mm} (B \to F \nu \bar \nu)_{\rm SM} } -1 \\[2mm]
& = -\frac{s_w^2}{X} \, \Delta C_9 + {\cal O} (M_W^4/M_{Z^\prime}^4) \,,
\end{split}
\end{equation}
with $X \simeq 1.47$ \cite{Brod:2010hi} denoting the SM loop contributions to $b \to s \nu \bar \nu$ from the $Z$-penguin and electroweak-box diagrams. Inserting~(\ref{Eq9}) into~(\ref{Eq14})  hence implies that a future measurement of $\Delta_{\nu \bar \nu}$, which is not  fully unrealistic  at SuperKEKB, would show an excess of  $20- 30\%$  (see also~\cite{Altmannshofer:2013foa}), if the $B \to K^\ast \mu^+ \mu^-$ anomaly and its implications survive further scrutiny. Notice that~(\ref{Eq14}) is a rather model-independent result, because it  only assumes lepton-flavour universality and the absence (or smallness) of right-handed currents. 

\subsubsection{Conclusions and outlook}

We have pointed out that in minimal $Z^\prime$ models that aim at explaining the pattern of NP suggested by the $3.7\sigma$ anomaly recently observed in $B \to K^\ast \mu^+ \mu^-$, there exists a triple-correlation that connects  NP in the Wilson coefficient of the semileptonic vector operator ($\Delta C_9$) to a shift  in the mass difference of $B_s$--$\bar B_s$ mixing ($\Delta_{B_s}$) and a violation  of first-row CKM unitarity ($\Delta_{\rm CKM}$). In fact, we have shown that the triple-correlation can be cast into a simple, analytic formula that relates the three quantities in question, and depends only logarithmically on the NP scale. As a result, precision determinations of $\Delta_{B_s}$ and $\Delta_{\rm CKM}$ can in principle be used to probe and overconstrain the $B \to K^\ast \mu^+ \mu^-$ anomaly in minimal $Z^\prime$ scenarios. These tests become more powerful for heavier $Z^\prime$ bosons, since $\Delta_{\rm CKM}$ being a one-loop effect  introduces a logarithm $\ln \hspace{0.25mm} (M_{Z^\prime}^2/M_W^2)$ that suppresses $\Delta C_9$ when written in terms of $\Delta_{B_s}$ and $\Delta_{\rm CKM}$. We emphasised that in models with purely vector couplings between the $Z^\prime$ and muons, the $B \to K^\ast \mu^+ \mu^-$ anomaly would leave no imprint in $B_s \to \mu^+ \mu^-$, which is only sensitive to the axial-vector part of the $Z^\prime$-muon coupling.  Due to $SU(2)_L$ invariance, effects in the $b \to s \nu \bar \nu$ transitions are on the other hand  inevitable, and amount to rate enhancements of around 25\%, in the case of theories with lepton-flavour universality and purely left-handed currents. 

In this note we have studied a toy model that contains only the $Z^\prime$-fermion couplings that are necessary to generate a large, negative contribution $\Delta C_9$. Clearly, such a model is not realistic  in the sense that in explicit $Z^\prime$ scenarios neither the axial-vector $Z^\prime$-muon coupling nor the flavour-diagonal $Z^\prime$-quark couplings will be exactly zero. Interestingly, CKM unitarity can be shown to  remain a stringent constraint on the structure of complete $Z^\prime$ models that can accommodate large shifts in $\Delta C_9$. Further powerful constraints on such scenarios   also arise from precision measurements of atomic parity violation and electron-electron M\o{}ller scattering, which are sensitive to the axial-vector $Z^\prime$-electron coupling. A detailed discussion of the possible phenomenological implications of the  $B \to K^\ast \mu^+ \mu^-$ anomaly will be presented in~\cite{inprep}. In fully realistic $Z^\prime$ models the triple-correlation found in our work will therefore not appear in its pure form.  On general grounds, certain correlations between the quark-flavour-changing $b \to s \ell^+\ell^-, \nu \bar \nu$ transitions and modifications in $\mu^- \to e^- \nu_\mu \bar \nu_e$ as well as  parity-violating $e^- \to e^-$ observables are however expected to remain, if a new neutral gauge boson should be responsible for the  deviations in $B \to K^\ast \mu^+ \mu^-$ as seen by LHCb. \\

\begin{acknowledgements}
We are grateful to Wolfgang Altmannshofer, Sebastian~J\"ager, Quim Matias and David Straub  for their valuable comments on the manuscript that allowed us to improve it. The research of R.~G. is supported by an STFC Postgraduate Studentship and F.~G. acknowledges support by the Swiss National Foundation under contract SNF 200021-143781.
\end{acknowledgements}

\end{document}